\documentclass[twocolumn,prd,aps,superscriptaddress,preprintnumbers,showpacs,nofootinbib,eqsecnum,amsfonts,amsmath]{revtex4-2}

\usepackage[english]{babel}
\usepackage{amsmath}
\usepackage{graphicx}
\usepackage{xspace}
\usepackage[colorlinks=true, allcolors=blue]{hyperref}

\newcommand{\ba}{\textsc{Bayestar}\xspace}
\newcommand{\pycbc}{PyCBC\xspace}
\newcommand{\pcl}{PyCBC Live\xspace}
\newcommand{\gstlal}{GstLAL\xspace}

\begin{document}

\title{Optimizing the low-latency localization of gravitational waves}

\author{Pierre-Alexandre \surname{Duverne}}
\affiliation{Universit\'e Paris-Saclay, CNRS/IN2P3, IJCLab, 91405 Orsay, France}
\affiliation{Universit\'e Paris Cit\'e, CNRS, Astroparticule et Cosmologie, F-75013 Paris, France}
\email{duverne@apc.in2p3.fr}

\author{Stéphanie \surname{Hoang}}
\affiliation{Universit\'e Paris-Saclay, CNRS/IN2P3, IJCLab, 91405 Orsay, France}

\author{Tito \surname{Dal Canton}}
\affiliation{Universit\'e Paris-Saclay, CNRS/IN2P3, IJCLab, 91405 Orsay, France}

\author{Sarah \surname{Antier}} 
\affiliation{Artemis, Observatoire de la C\^ote d'Azur, Universit\'e C\^ote d'Azur, Boulevard de l'Observatoire, 06304 Nice, France}

\author{Nicolas \surname{Arnaud}} 
\affiliation{Universit\'e Paris-Saclay, CNRS/IN2P3, IJCLab, 91405 Orsay, France}
\affiliation{European Gravitational Observatory (EGO), I-56021 Cascina, Pisa, Italy}

\author{Patrice \surname{Hello}}
\affiliation{Universit\'e Paris-Saclay, CNRS/IN2P3, IJCLab, 91405 Orsay, France}

\author{Francesco Pannarale}
\affiliation{Dipartimento di Fisica, Università di Roma ``Sapienza'', Piazzale A. Moro 5, I-00185, Roma, Italy}
\affiliation{INFN Sezione di Roma, Piazzale A. Moro 5, I-00185, Roma, Italy}

\date{\today}

\begin{abstract}
Gravitational-wave data from interferometric detectors like LIGO, Virgo and KAGRA is routinely analyzed by rapid matched-filtering algorithms to detect compact binary merger events and rapidly infer their spatial position, which enables the discovery of associated non-GW transients like GRB 170817A and AT2017gfo.
One of the critical requirements for finding such counterparts is that the rapidly inferred sky location, usually performed by the \ba algorithm, is correct.
The reliability of this data product relies on various assumptions and a tuning parameter in \ba, which we investigate in this paper in the context of \pcl, one of the rapid search algorithms used by LIGO, Virgo and KAGRA.
We perform simulations of compact binary coalescence signals recovered by \pcl and localized by \ba, under various configurations with different balances between simplicity and realism, and we test the resulting sky localizations for consistency based on the widely-used PP plot.
We identify some aspects of the search configuration which drive the optimal setting of \ba's tuning parameter, in particular the properties of the template bank used for matched filtering.
On the other hand, we find that this parameter does not depend strongly on the nonstationary and non-Gaussian properties of the detector noise.
\end{abstract}

\maketitle

\section{Introduction}

The rapid detection of gravitational-wave (GW) signals from compact binary coalescences (CBCs) in data from existing observatories (LIGO \cite{ligo}, Virgo \cite{virgo} and KAGRA \cite{kagra}) is done by running dedicated low-latency searches such as \pcl \cite{pycbc_live_1, pycbc_live_2}, \gstlal \cite{GstLAL}, MBTAOnline \cite{mbta}, SPIIR \cite{SPIIR} or cWB online \cite{CWB}. These algorithms enable the production of rapid GW alerts usable by other instruments to search for counterparts. During the O3 observing run \cite{GWTC-3,PhysRevX.11.021053}, that happened between April 2019 and March 2020, these alerts were publicly distributed with notices and circulars sent via the General Coordinates Network (GCN) \cite{GCN}, as it is the case during the current O4 run.
Once low-latency search pipelines identify a candidate CBC signal with a sufficiently low false-alarm rate, the spatial location of the event is estimated by a dedicated algorithm called \ba \cite{bayestar}, which uses a Bayesian framework to rapidly infer the joint posterior distribution for the source's sky location (right ascension $\alpha$ and declination $\delta$) and the luminosity distance.
Other telescopes, or telescope networks, can then use these inferred quantities to follow up GW alerts.
At longer time scales, typically at least tens of minutes after the merger time, a full parameter inference is performed on the detected signals \cite{morisaki2023rapid, Ashton_2019}, which can be used to update the \ba result.

The rapid distribution of this information has played a key role in the discovery of the GW170817 event, the first detection of a binary neutron star merger (BNS) associated with the short gamma-ray burst (GRB) GRB\ 170817A and the optical counterpart (kilonova) AT2017gfo less than 11 hours after the merger \cite{grb_2017, Goldstein_2017, mma170817, D_Avanzo_2018, Alexander_2017, Hallinan_2017, Troja_2017, Soares_Santos_2017}. For this particular event, the 90\% credible area of the sky location was about $28$ deg$^2$. In general, however, this area can be much larger, depending on the signal-to-noise ratio (SNR) of the GW signal and on the status of the detector network at the time of the signal (one, two or three interferometers observing). For instance, the median size of the 90\% credible region was about $1000$ deg$^2$ for Binary Black Hole (BBH) alerts during the O3 run, and predictions made before the O4 run give similar orders of magnitude ($>1000$ deg$^2$) \cite{petrov2022, Kiendrebeogo_2023}.

Since the success of follow-up campaigns depends on the availability of prompt sky localizations of GW events, confidence in \ba results is crucial, especially for 1) well-localised events and 2) events containing at least one neutron star \textit{à la GW170817}.
\ba's rapidity (typically $\sim 1$ s) rests on a simplified treatment of the full parameter-estimation problem, in particular on the decoupling of the extrinsic parameters describing a CBC signal (spatial location and orientation) and its intrinsic parameters (typically masses and spins). Although this simplification has proven to be extremely useful, in principle, it could be associated with biases in the spatial localization.

For instance, \ba's self-consistency has been tested for candidates produced by the \gstlal algorithm \cite{bayestar} and revealed a systematic underestimation of the credible area of the sky location. This underestimation was corrected by introducing a correction parameter that we denote $\xi$ for this work. The optimal value for $\xi$ was then found using a signal simulation campaign and the \gstlal pipeline, resulting in $0.83$, which is used to date irrespective of which search pipeline produced the candidate. On the other hand, \cite{pycbc_live_1} pointed out a possible overestimation of the credible area for candidates reported by the PyCBC Live pipeline, raising the question of whether the optimal value of $\xi$ might depend on the details of how each GW candidate is generated. Overestimating the credible area is not a priori problematic, as the GW source is still inside the sky area covered by the telescopes. However, it makes the electromagnetic follow-up more demanding, wasting telescope time. An underestimation of the uncertainty in the sky location is even less desirable, as the follow-up risks missing the counterpart if the GW source is outside of the high-probability region.

Two points were not directly addressed in \cite{bayestar}. First, what specific aspect(s) of the entire detection and inference process introduce the need for such a correction? \cite{bayestar} lists a number of plausible possibilities, for instance, certain properties of the template waveforms used for matched-filtering, or the basic approximation made by \ba. Second, does the optimal value depend on which search pipeline produced the candidate, or is $0.83$ sufficient for all the existing search pipelines? The latter question is further raised by the PP-plot in Figure 5 of \cite{pycbc_live_1}, which was constructed using a \pcl analysis, and shows a curve above the expected diagonal, consistent with an overestimation of the uncertainty.

Motivated by these considerations, in this paper we present a detailed exploration of possible sources of bias in the sky localization produced by \ba, focusing on CBC candidates produced by the \pcl pipeline and on the behavior of the $\xi$ parameter described in \cite{bayestar}.

\section{Low latency GW search and localization}
\label{sec:ll-search}

The most sensitive search algorithms for GWs from stellar-mass CBCs are based on a technique called \emph{matched filtering} that consists in cross-correlating the data with a CBC template waveform to compute the signal-to-noise ratio.
The template is taken from a pre-computed collection of CBC waveforms called a template bank.
The bank is designed to efficiently cover the parameter space of the binary's four intrinsic parameters (the masses and spins of the two merging objects) under various simplifying assumptions, namely that the binary orbit is quasi-circular, that the spins are aligned with the orbital angular momentum, that matter effects do not significantly distort the waveform, and that the SNR is dominated by the quadrupolar mode of the radiation.

Although there are various implementations of matched filtering, for this work, we focus on the algorithm called \pcl \cite{pycbc_live_1, pycbc_live_2}, where matched filtering is computed in the frequency domain.
It is based on the \texttt{FindChirp} algorithm \cite{findchirp}, where the GW data $s$ are correlated with a template waveform $h$ in the frequency domain to build the complex SNR time series:
\begin{equation}
    \label{eq:snr}
    z(t) = \frac{4}{\sqrt{\langle h | h \rangle}} \int^{\infty}_{0} \frac{\Tilde{s}(f)\Tilde{h}^{*}(f)}{S_{n}(f)} e^{2\pi i f t}\,df
\end{equation}
where 
$f$ denotes the value of the Fourier frequency
and the $\langle .|. \rangle$ operator corresponds to the inner product 
        \begin{equation}
        \label{eq:sca}
            \langle a|b \rangle = 4 \int^{\infty}_{0} \frac{\Tilde{a}(f)\Tilde{b}^{*}(f)}{S_{n}(f)}\,df.
        \end{equation}
The normalisation $S_{n}(f)$ corresponds to the one-sided power spectral density (PSD) of the noise around time $t$.
In \pcl, the PSD is estimated using the Welch method \cite{welch} with a median averaging.
A trigger is identified in the data from a single detector whenever a local maximum in the magnitude of the SNR time series $\rho(t) = |z(t)|$ crosses a threshold of $4.5$. The time and template associated with the local maximum are recorded as part of the trigger.
\pcl can then use the set of triggers from multiple detectors to either identify single-detector candidates (signals detectable in a single detector only) or coincident candidates (signals detectable in at least two detectors). A significance (false-alarm rate, FAR) is calculated in different ways in the two cases. Candidates with a FAR lower than one per two hours are transmitted to the Gravitational-wave Candidate Event DataBase (GraceDB) \cite{gracedb}.

\ba \cite{bayestar} is then used to infer the spatial location of each candidate uploaded to GraceDB, in terms of the posterior probability for the three-dimensional position of the source in the universe, under a Bayesian framework.
\ba takes as input the point estimates of the masses and spins from the template reported by \pcl and associated with the candidate, a portion of the complex SNR time series around the estimated merger time, and the sensitivities of each involved detector at the time of the candidate. Based on this information, \ba computes the likelihood and infers the source position. Using the point estimates of masses and spins is an approximation that avoids having to jointly infer the spatial location and the intrinsic parameters of the binary (component masses and spins), making the algorithm rapid.
The resulting spatial posterior is produced within seconds, enabling rapid electromagnetic follow-up of the candidate.
The approximation is justified because the uncertainties in the intrinsic and extrinsic parameters are typically very weakly correlated; this can be understood intuitively considering that the spatial location is mainly determined by the relative arrival times, phases and amplitudes at the different detectors, while mass and (aligned) spin parameters are mainly determined from the temporal evolution of the GW waveform.

For this work, we focus on the two-dimensional posterior probability obtained by marginalizing the three-dimensional one over distance, which is the quantity of immediate interest for deciding where to point telescopes.
We will refer to this quantity as the \emph{skymap} for simplicity.

\section{Motivation and Methodology}
\label{sec:self}

A common way to test the reliability of parameter inference in GW astronomy is to use a test based on the Percentile-Percentile plot (PP-plot) \cite{PhysRevD.89.084060, bayestar} and this is what we use in this study.
We simulate a population of CBC signals (\emph{injections}), add them to noise (either simulated or real, as discussed case-by-case later), produce a candidate detection for each signal, and use \ba to produce the skymaps for each candidate. We then perform a cumulative sum of the probability contained in the skymap pixels, starting from the highest-probability pixel and proceeding in order of descending probability, and stopping at the pixel that contains the true position. The cumulative sum corresponds to the cumulative probability in the smallest area that contains the true location, and is referred to as the \emph{search probability} in the following. Repeating this procedure for the whole population of simulated CBC signals, we then make a cumulative histogram of the search probability. This histogram is called a percentile-percentile plot and allows one to evaluate whether the localization is self-consistent. In the case of a consistent localization, the histogram is expected to be diagonal, representing the fact that the search probability distribution is uniform. In other words, one should find $p$\% of the simulated signals to have their true location within the $p$\% credible region.

In the following sections, we will perform a sequence of simulations to test some of the possible origins of the correction introduced via the $\xi$ parameter, starting with the simplest controlled simulations and progressing towards full end-to-end simulations with a realistic configuration of \pcl.

\subsection{Choices of signals and detectors}
\label{sec:pop}

\begin{figure}
    \includegraphics[width=\columnwidth]{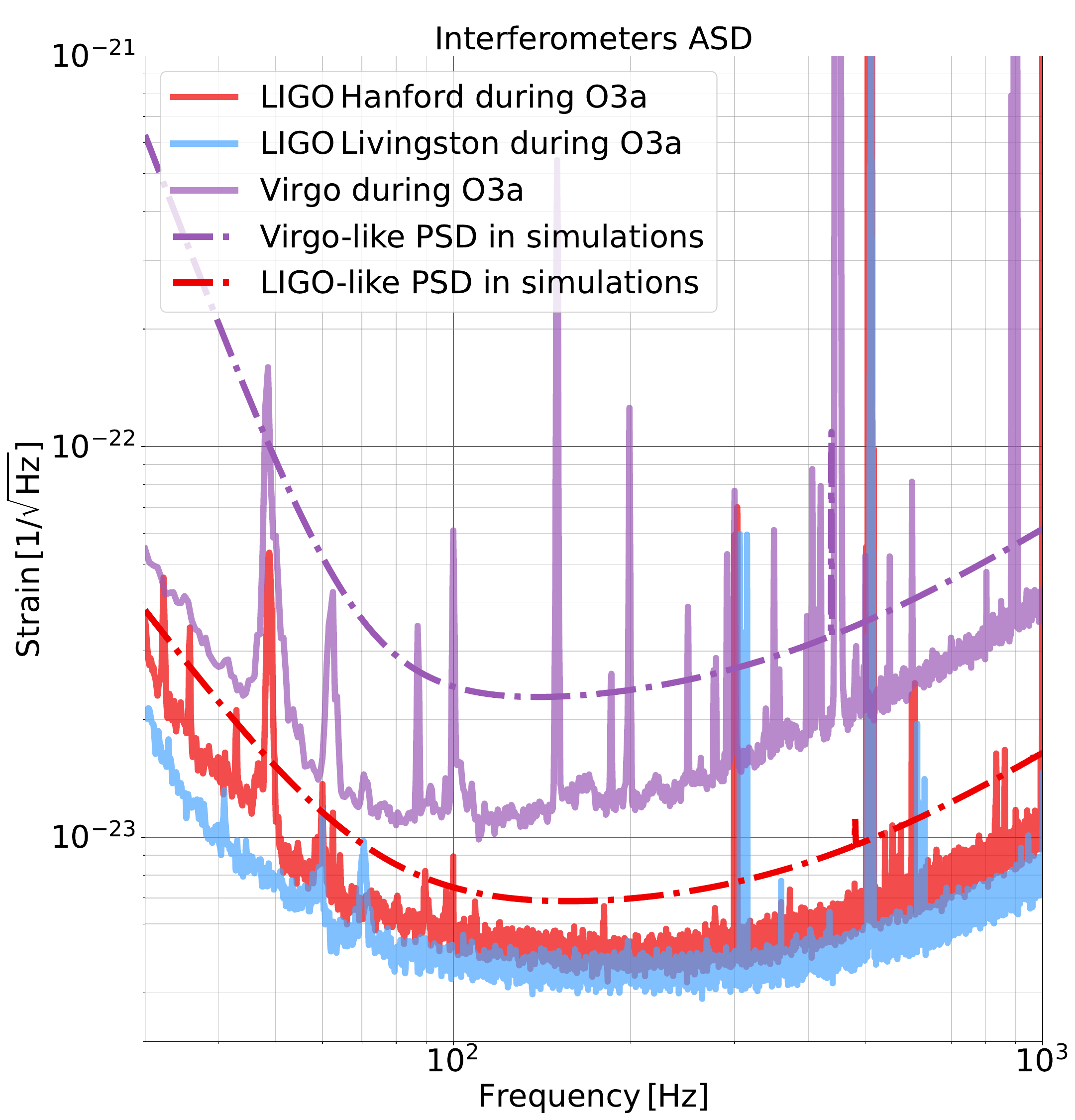}
    \caption{Noise ASD models used to generate the simulated data (dash-dotted lines). These are specifically \texttt{aLIGOMidLowSensitivityP1200087} and \texttt{AdVEarlyLowSensitivityP1200087} from the LALSimulation library \cite{lalsuite}. For comparison, we also show ASDs representing the three detectors’ strain sensitivities during O3, taken from the GWTC-2 catalog \cite{PhysRevX.11.021053}.}
    \label{fig:psd}
\end{figure}

The choices of the parameters of the simulated CBC signals are summarised in Table~\ref{tab:injections}. For BNS signals, we use inspiral-only analytical waveforms based on the \texttt{SpinTaylorT4} approximant \cite{PhysRevD.80.084043}. For BBH and NSBH (neutron-star-black-hole) signals we use effective-one-body (EOB) inspiral-merger-ringdown waveforms based on the \texttt{SEOBNRv4} approximant \cite{PhysRevD.95.044028}.
For all signals, the spins of the coalescing objects are assumed to be aligned with the orbital angular momentum.
\begin{table}
    \begin{tabular}{cc}
        Parameter & Distribution \\
        \hline
        \hline
        Polarisation angle & $\mathcal{U}([0, 2\pi])$ \\
        Phase at coalescence  & $\mathcal{U}([0, 2\pi])$ \\
        $\cos{\iota}$  & $\mathcal{U}([-1, 1])$ \\
        BNS position  & $\mathcal{U}(\mathbb{S}(0, [40, 250]$Mpc))  \\
        NS mass  & $\mathcal{N}(\mu=1.4$M$_{\odot},\,\sigma=0.1$M$_{\odot})$ \\
        Aligned comp.~of NS spin  & $\mathcal{N}(\mu=0,\,\sigma=0.01)$ \\
        BBH position  & $\mathcal{U}(\mathbb{S}(0, [100, 5000]$Mpc))  \\
        BH mass  & $\mathcal{U}([5, 100]$M$_{\odot}$) \\
        Aligned comp.~of BH spin  & $\mathcal{N}(\mu=0,\,\sigma=0.1)$ \\
        \hline
    \end{tabular}
    \caption{
        Parameters of the simulated CBC signals. $\iota$ is the orbital inclination of the binary. The notation $\mathcal{U}([a, b])$ indicates a uniform distribution over the interval $[a, b]$. $\mathcal{N}(\mu,\,\sigma)$ indicates a normal distribution with mean $\mu$ and standard deviation $\sigma$.
        The $\mathcal{U}(\mathbb{S}(0, [a, b]$)) notation corresponds to a uniform distribution in volume in a shell centered on the origin, with inner radius $a$ and outer radius $b$. Masses are expressed in the detector frame, i.e.~redshifted.
    }
    \label{tab:injections}
\end{table}
The simulated waveforms are injected in 600 s data segments of Gaussian and stationary noise, with the exception of Section~\ref{sec:real} where we will use real noise. The simulated noise has amplitude spectral densities (ASDs) shown in Figure~\ref{fig:psd}. We use the same ASD for the two LIGO interferometers and a significantly less sensitive ASD for Virgo, so as to have a network configuration with relative sensitivities similar to O3 and O4.
The simulated interferometers are overall less sensitive than the O3 instruments, particularly Virgo in the low-frequency region. However, this is not expected to significantly impact our conclusions about the localization consistency.

In order to simulate the selection effects of CBC search pipelines, we apply a cut on the optimal SNR of each signal before generating the skymap. 
The optimal SNR corresponds to the expected SNR a GW signal would have if the template was identical to the signal.
The optimal SNRs of the individual detectors involved in a GW detection are summed in quadrature to compute the network's optimal SNR.
We keep only the signals with a network SNR above 8 and with at least one individual optimal SNR above $5.5$. These cuts are chosen to match Table 1 from \cite{GWTC-3}, which summarises the network SNR of the publicly distributed alerts during O3, and SNR $\sim 8$ is the lower bound for distributed online GW candidate events. The signals surviving these cuts are localised with \ba to generate the skymaps.

\subsection{Baseline percentile-percentile test}
\label{sec:pp-test}

As a first check, we attempt to reproduce the results in Figure 5 of \cite{pycbc_live_1} using 500 BNS, 500 NSBH and 500 BBH simulated following the distributions in Section~\ref{sec:pop}.

For our first simulations, we use a tool available in \pycbc, called \texttt{pycbc\_make\_skymap}\footnote{\label{foot:skm} \url{https://github.com/gwastro/pycbc/blob/master/bin/pycbc_make_skymap}}, which simulates the PyCBC Live search pipeline in a controlled, faster and simplified way. This tool creates a template waveform from a user-provided set of parameters, performs the matched filtering on the simulated data, calculates the SNR time series, formats the results and passes them to \ba to produce the skymap.
An important simplification is that \texttt{pycbc\_make\_skymap} uses a single template specified by the user, thus removing the possible effects related to a template bank.
After running \texttt{pycbc\_make\_skymap} and \ba, we calculate the search probability for each signal and use its distribution to obtain the PP plot.

Apart from showing the PP plot, we also perform a Kolmogorov-Smirnov (KS) test \cite{ks_test} as a quantitative check of the consistency of the localization.
The null hypothesis of our KS test is that the localization is self-consistent.
As described in Section~\ref{sec:self}, under this hypothesis, the search probability is uniformly distributed in $[0,1]$ and produces a diagonal cumulative distribution. Correspondingly, we use the KS-test to quantify the agreement of the search probability distribution with a uniform distribution. In the following sections, we consider that a KS-test p-value smaller than $3\times 10^{-3}$, corresponding to a $\sim$3-$\sigma$ deviation, is small enough to reject the null hypothesis.

\begin{figure}
    \includegraphics[width=\columnwidth]{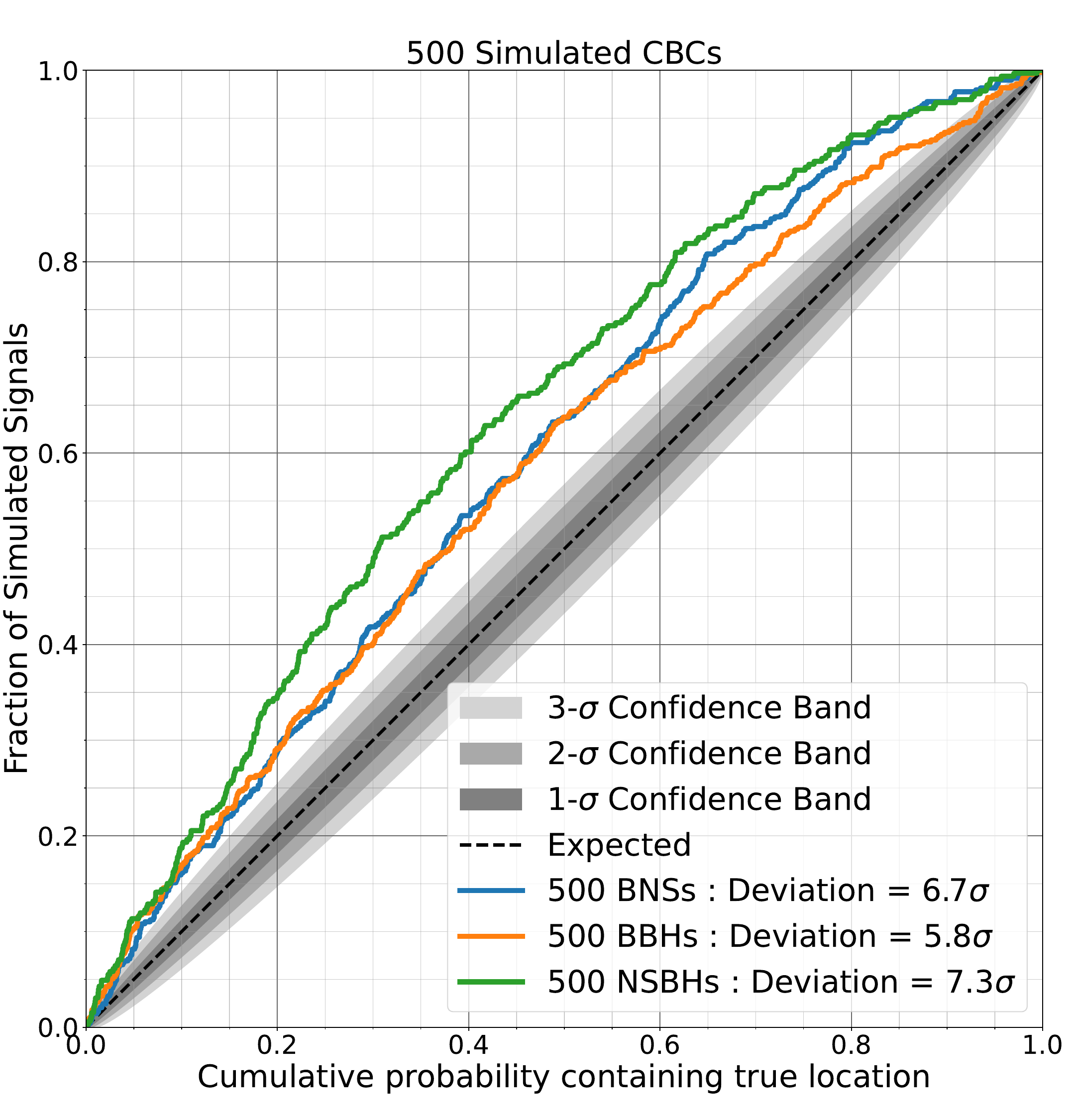}
    \caption{Baseline PP-plot for the simulated BNSs, 
    NSBHs and BBHs. The blue curve is the cumulative distribution of the search probability computed with the skymaps of 500 BNSs. The orange one is for the 500 simulated BBHs. The green one is for the 500 simulated NSBHs. 
    In the case of a self-consistent analysis, the curve is expected to be diagonal. The grey-shaded areas around the diagonal are the 1, 2 and $3\sigma$ confidence bands, showing the variability expected from the finite number of simulated signals.}
    \label{fig:pp-plot-init}
\end{figure}

For the first set of simulations, we use \ba in its O3 configuration and assume that the true intrinsic parameters (masses and spins) of the signals are exactly known, i.e.~we use the true parameters of each signal in \texttt{pycbc\_make\_skymap}. The baseline PP plot for this configuration is visible in Figure~\ref{fig:pp-plot-init} with BBHs in orange, NSBHs in green
and BNSs in blue. For all three source types, the curves are clearly inconsistent with the null hypothesis, and we conclude that the sky-localization uncertainties are overestimated, which appears to confirm the results in \cite{pycbc_live_1}.

\subsection{Influence of the network configuration}
\label{sec:network-config}

   During an observing run of the LIGO and Virgo interferometers, the number of observing instruments varies on a time scale of hours for various reasons, including maintenance, earthquakes, and local technical issues. Therefore, the sky localizations of different events from the same run, even within a few hours of each other, may be performed with different detector networks. Figure 5 of \cite{pycbc_live_1} showed a PP-plot where the deviation is larger when the network includes Virgo (green curve) than when only LIGO-Hanford and LIGO-Livingston are used (orange curve). In order to better understand this effect, we simulated two-interferometer network configurations with the same sensitivities as in the previous Section to test the influence of the number of instruments on the accuracy of the sky localization. We used 500 BBH injections for each possible pair of instruments: H1-L1, L1-V1 and H1-V1.
    In addition, since Virgo is much less sensitive than the LIGO detectors, we also explored the influence of the relative sensitivities by considering a three-detector network where all detectors have the same LIGO sensitivity given by the dash-dotted red curve in Figure~\ref{fig:psd}.

   The results are summarized in Table \ref{tab:net-conf}. All the p-values returned by the KS-test allow us to reject the null hypothesis, hence the deviation remains present even with network configurations that are significantly different than our baseline. The p-value for the H1-L1 configuration is much larger than for the others (i.e.~the deviation is smaller) and the three-detector configurations considered so far exhibit larger deviations than two-detector ones. Note that the H1-L1 configuration is the most geometrically degenerate one, since the detectors are spatially closer and their antenna pattern functions are oriented almost in the same way.

    In light of these results, and of the fact that additional detectors like KAGRA and LIGO-India \cite{LIGOIndia} are planned to join in the future, in the following sections we will no longer consider two-detector networks, and rather restrict our attention to our baseline three-detector configuration with the sensitivities presented in Figure~\ref{fig:psd}.

    \begin{table}    
    \begin{center}
    \begin{tabular}{ccc}
    \hline
    \textbf{Network configuration } & \textbf{p-value} & \textbf{Deviation}\\
    \hline
        H1-L1 & $4.1 \times 10^{-4}$ & $3.5\sigma$\\
        H1-V1 & $1.3 \times 10^{-8}$ & $5.7\sigma$\\
        L1-V1 & $4.0 \times 10^{-7}$ & $5.1\sigma$\\
    \hline
        H1-L1-V1 with LIGO PSD & $ 1.0 \times 10^{-10}$ & $6.4\sigma$\\
    \hline
    \end{tabular}
    \caption{Results of the KS test for different two-interferometer configurations with the same sensitivities as in Section~\ref{sec:pp-test}, and the three-interferometer configuration with identical sensitivities for all interferometers.}
    \label{tab:net-conf}
    \end{center}
    \end{table}
    
\section{Origin of the deviation}
\label{sec:origin}

\begin{figure}
        \includegraphics[width=\columnwidth]{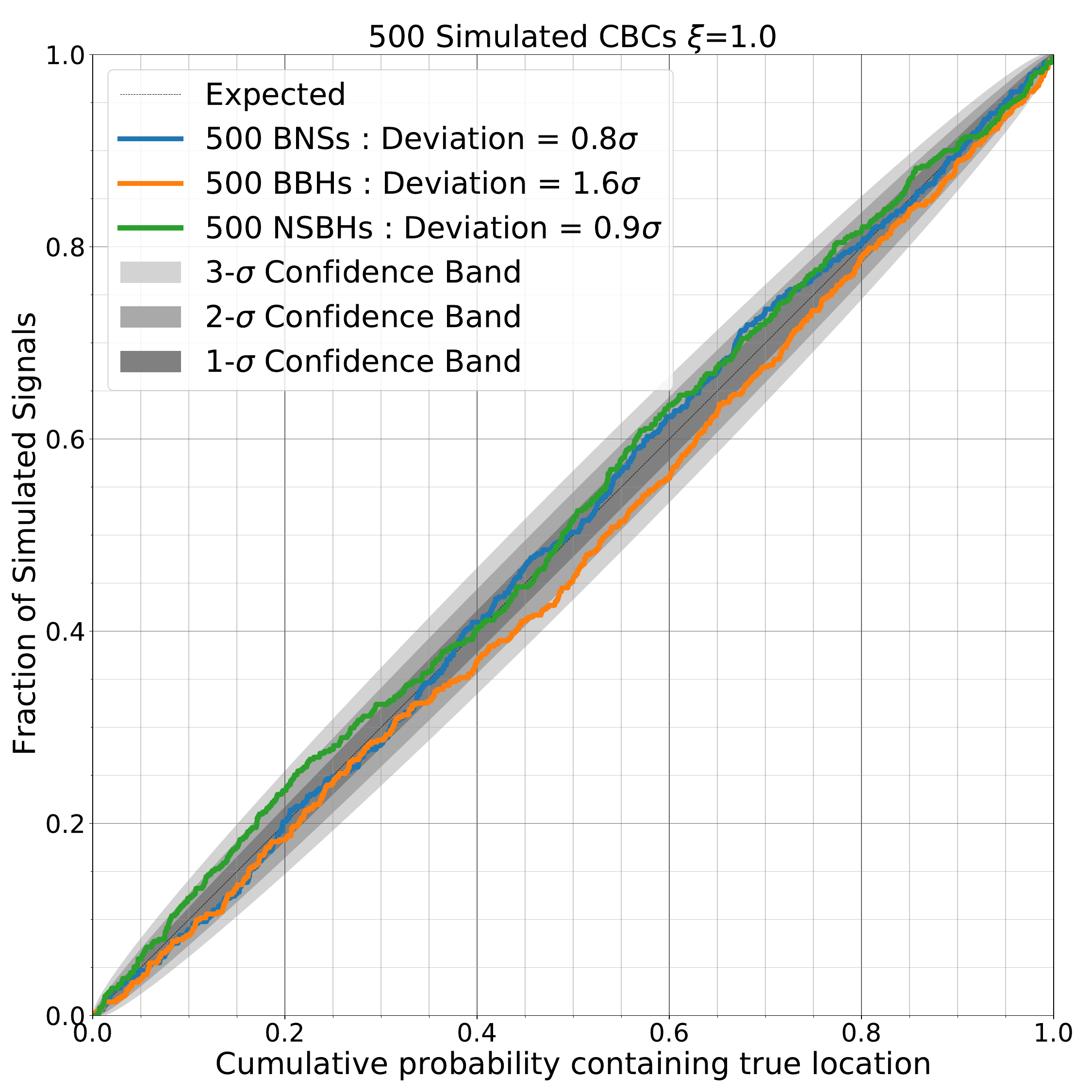}
        \caption{PP plot for three separate populations of GW sources after setting the $\xi$ parameter to 1. All the deviations are within 3-$\sigma$. Consequently, we cannot reject the null hypothesis.}
        \label{fig:pp-plot-ff}
\end{figure}

As introduced in the previous Sections, \ba uses a factor $\xi$ to ensure that the PP-plot is diagonal, and its default value $0.83$ has been found via simulations recovered by the \gstlal pipeline.
Our next variation with respect to the baseline result is removing the correction introduced via this factor.
We set $\xi=1$ and repeat our simulations, using three separate populations of $500$ BBH, $500$ BNS and $500$ NSBH mergers to test whether the effect of $\xi$ could depend on the source type.
The results for these simulations are summarised in Table~\ref{tab:ff}, and the PP plots are visible in Figure~\ref{fig:pp-plot-ff}. 
For those three simulated populations, the p-values returned by the KS-test are not small enough to reject the null hypothesis, i.e.~the \ba~analysis is self-consistent. Hence, we conclude that the deviation observed in Figure \ref{fig:pp-plot-init} can be explained by the default value of $\xi$ being different from $1$, and it appears that our simulations are localized in a perfectly consistent way without the need for the $\xi$ correction.

\begin{table}
    \begin{center}
        \begin{tabular}{ccc}
        \hline
        \hline
        \textbf{CBC Type} & \textbf{p-value} & \textbf{Deviation}\\
        \hline
        BNS  & $0.45$ & $0.76\sigma$\\
        BBH  & $0.11$ & $1.6\sigma$\\
        NSBH & $0.36$ & $0.92\sigma$\\
        \hline
        \end{tabular}
        \caption{Results of the KS-test for our second set of simulations, testing the effect of setting $\xi = 1$.}
        \label{tab:ff}
    \end{center}
\end{table}

After this second test, we are thus left with a number of questions. What is different between the simulation campaign done in \cite{bayestar} and our simulations? Do the different values of $\xi$ depend on the fact that we assumed idealized noise, or on differences between the \gstlal and \pcl analyses, or on the simplifications made by \texttt{pycbc\_make\_skymap}?
These questions are investigated in the following sections.

\section{Effect of real detector noise}
\label{sec:real}

In the first sets of injections, the signals have been injected in simulated Gaussian and stationary noise. However, the data produced by a real GW detector have a non-stationary character and contain instrumental transients, commonly called glitches, that may have an influence on the localization (see e.g.~GW170817, where a loud glitch in LIGO Livingston's data had to be removed before any analysis \cite{GW170817Discovery}).
Thus, it is necessary to verify if these effects might be causing a bias in the localization, introducing the need for the $\xi$ correction.
We simulate populations of BBH, NSBH and BNS in the same way as described in Section \ref{sec:pop} and inject the signals into publicly available data from O3a (April to October 2019) and O3b (November 2019 to March 2020).
The injections are made into data segments of O3 by picking a random time and keeping it if $600$ s of data were available in the three detectors around that time.
As sensitivity and glitch rate significantly improved between the two halves of O3, we created two separate sets of injections for O3a and O3b.
    
\begin{figure*}
        \includegraphics[width=\textwidth]{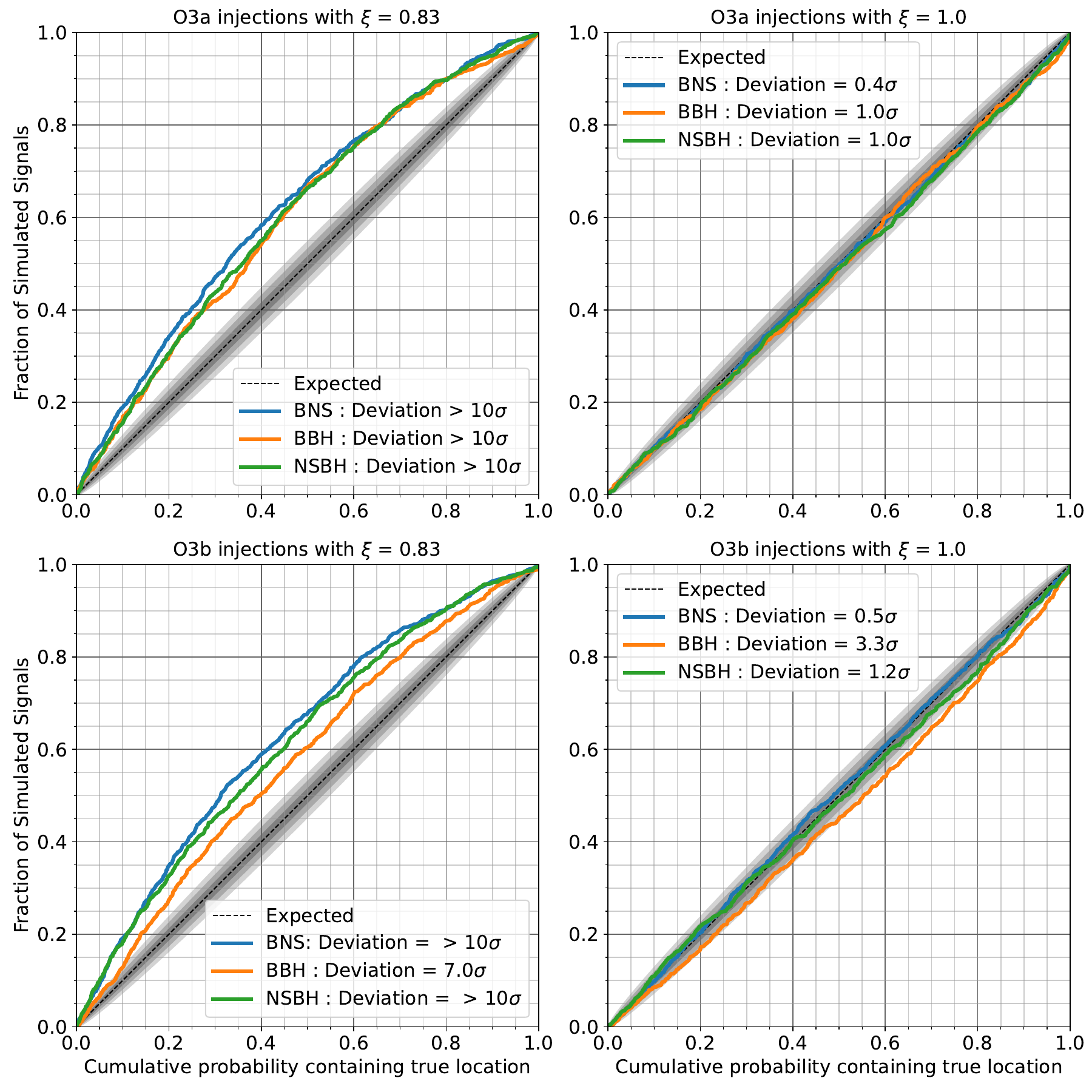}
        \caption{
        PP-plots for injections in real noise for O3a and O3b. The upper row corresponds to injections in O3a data, and the bottom one to the O3b injections. The left column corresponds to localizations made with $\xi=0.83$ and the right column to localizations with $\xi=1$. In each plot, the orange curve is for the BBH, the blue for BNS and the green curve is for NSBH.
        }
        \label{fig:o3b}
\end{figure*}

We inject 1000 BBH signals, 1000 NSBH and 1000 BNS for both O3a and O3b data segments, and we keep working with the assumption that the intrinsic parameters are exactly known, as in Section~\ref{sec:pp-test}. For each interferometer, an average ASD is estimated for each week of the O3 run. Then for each signal, the closest ASD  to the picked time is chosen for the injection. Eventually, 908 BBHs, 888 NSBHs and 921 BNSs are loud enough to be detected and localised with O3a data. For O3b, we have 912 BBHs, 879 NSBHs and 940 BNSs. The results are summarised in Table \ref{tab:ff-reala}, and Figure~\ref{fig:o3b} shows the PP plots for the injected signals.

For both O3a and O3b, we observe deviations similar to Figure~\ref{fig:pp-plot-ff} for $\xi=0.83$, and much smaller deviations for $\xi=1$. 
The case of the BBH injected in O3b data shows a deviation of $3.3 \sigma$ for $\xi=1$, indicating a small but statistically significant deviation based on the threshold we decided in Section~\ref{sec:pp-test}. We can see, indeed, that the BBH curve sags below the diagonal with $\xi=1$, i.e.~the uncertainty is slightly underestimated. This is not the case for BNS or NSBH signals. This might indicate a small mass dependence on how real detector noise affects the sky localization. Indeed, a recent study, which investigates the effect of specific glitch families on sky localizations produced via \pcl and \ba, also draws a similar conclusion \cite{Macas2022}.

This mass dependence, however, appears fairly small in our study, and only detectable with O3b data.
We conclude that setting $\xi=0.83$ does not appear to be required because of the real noise properties, like the presence of glitches in the data.
Consequently, we proceed with investigating the remaining questions using simulated Gaussian noise, which is easier to work with.

\begin{table}[h]
    \begin{tabular}{c|ccc|c|ccc}
    \textbf{Run} & \textbf{CBC} & \textbf{$\xi$} & \textbf{Deviation} & \textbf{Run} & \textbf{CBC} & \textbf{$\xi$} & \textbf{Deviation}\\
    \hline
    \hline
       & BBH  & $0.83$ & $\geq 10\sigma$ &      & BBH &  $0.83$ & $7.0\sigma$  \\
       &      & $1.0$  & $1.0\sigma$     &      &     &  $1.0$  & $3.3\sigma$ \\
  O3a  & BNS  & $0.83$ & $\geq 10\sigma$ &  O3b & BNS &  $0.83$ & $\geq 10\sigma$ \\
       &      & $1.0$  & $0.37\sigma$     &      &     &  $1.0$  & $0.54\sigma$ \\
       & NSBH & $0.83$ & $\geq 10\sigma$ &      & NSBH&  $0.83$ & $\geq 10\sigma$ \\
       &      & $1.0$  & $1.0\sigma$     &      &     &  $1.0$  & $1.2\sigma$ \\
    \hline
    \end{tabular}
    \caption{Results of the PP-plot and KS tests for the simulations testing glitches and non-stationarities in O3a data on the left and O3b on the right.}
    \label{tab:ff-reala}
\end{table}

\section{Effect of incorrect template parameters}
    \label{sec:blurr}
    As stated in \cite{bayestar}, one of the possible explanations for why $\xi$ is necessary might be the inevitable discrepancy between the true signals and the template waveforms reported with the detection candidates by the matched filtering algorithms.
    Such a discrepancy arises mainly from two effects: the detector noise and the various approximations made in the algorithms themselves, including the physics neglected in the models used for the search templates.
    
    In our simulations up to this point, we used the exact same waveform for both the injection and the localization, which implies assuming that the search would report exactly the true intrinsic parameters (masses and spins) of the sources. We now relax this assumption and perform additional simulations for which the intrinsic parameters of the template used for \ba are randomized in a way that mimics the effect of an online low-latency search. We focus on BNS signals only, for they are the most promising GW source for having EM counterparts, making their low-latency localization critical. In addition, most of the previous tests show qualitatively similar behavior between BNS, NSBH and BBH systems, as visible in Figures~\ref{fig:pp-plot-init}, \ref{fig:pp-plot-ff} and~\ref{fig:o3b}.

\begin{figure}
        \includegraphics[width=\columnwidth]{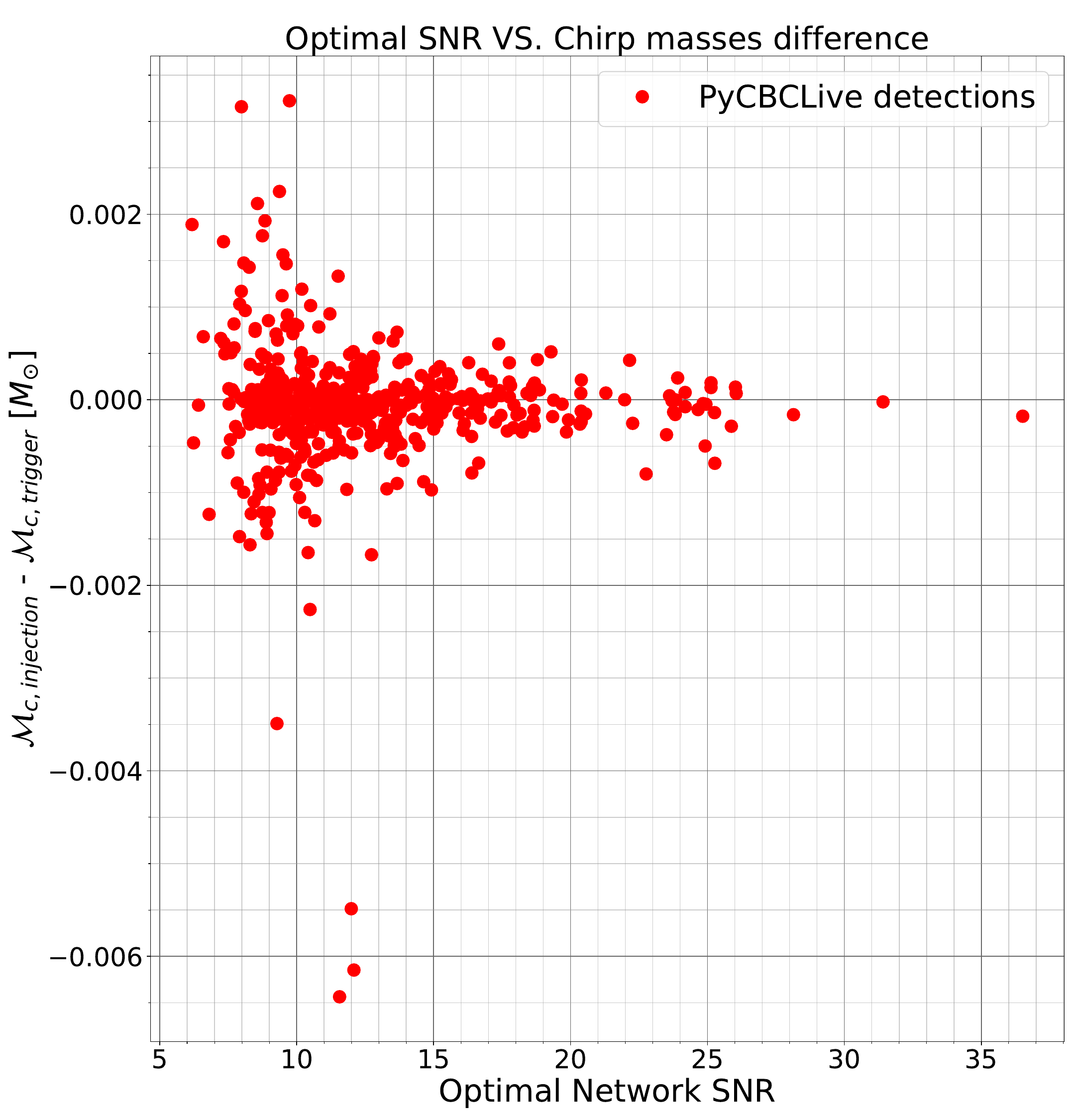}
        \caption{Difference between the true ($\mathcal{M}_{c,injection}$) and recovered $(\mathcal{M}_{c,trigger}$) redshifted chirp masses for a population of BNS signals recovered by an end-to-end search based on \pcl, as a function of the intrinsic SNR of each signal. These results are used to estimate the scatter in recovered chirp mass which affects the template used for \ba.}
        \label{fig:chirp-distrib}
    \end{figure}

\begin{figure}
        \includegraphics[width=\columnwidth]{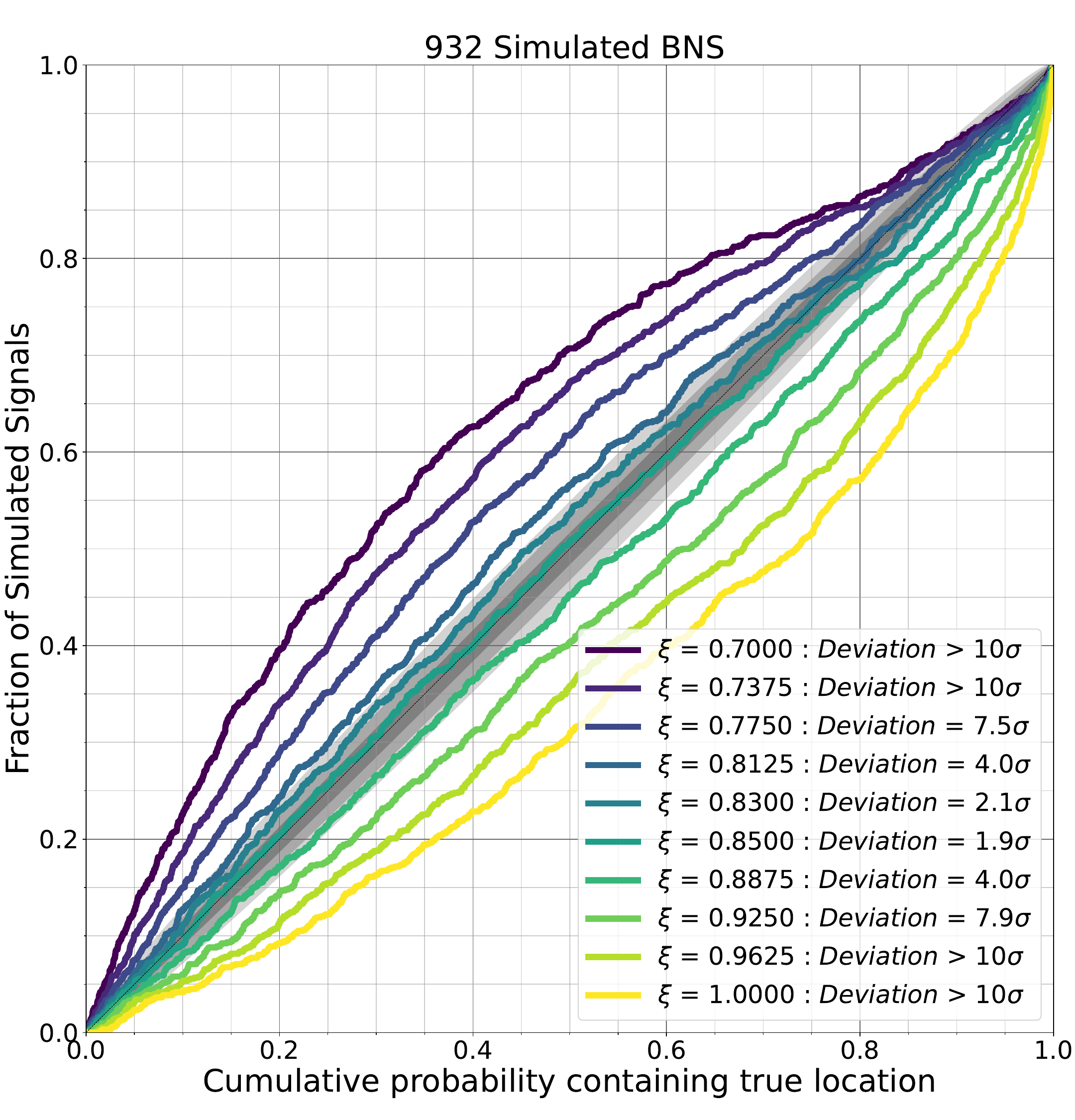}
        \caption{PP-plots obtained after randomising the template's intrinsic parameters compared to their true values before the signal localization.}
        \label{fig:pp-plot-all-blur}
    \end{figure}

    We inject simulated signals in 600 s segments of Gaussian noise. The aligned-spin parameters of the templates given to \ba are chosen uniformly in $[-1,1]$, to mimic the fact that search algorithms like \pcl do not typically provide reliable point estimates of the spins. The redshifted component masses $m_{1,2}$ of the templates are chosen by expressing them via the mass ratio
    \begin{equation}
        q = \frac{m_1}{m_2} \quad m_1 \geq m_2,
    \end{equation}
    and the chirp mass
    \begin{equation}
       \mathcal{M}_{c} = \frac{(m_1 m_2)^{3/5}}{(m_1 + m_2)^{1/5}}.  
    \end{equation}
    Similarly to spins, for the mass ratio $q$ we choose a uniformly distributed value in $[1, 3]$, which is consistent with the range of neutron star masses commonly assumed within LIGO and Virgo, i.e.~$[1, 3] M_{\odot}$. Dealing with the chirp mass $\mathcal{M}_{c}$ is more delicate, for this parameter is very well reconstructed by an online search, in particular for BNS signals \cite{Biscoveanu_2019}. To randomize the chirp mass in a realistic way, we use the results obtained with an end-to-end search performed with \pcl on BNS injections described later in Section~\ref{sec:ff-live}. The produced results are used to compute the standard deviation $\sigma_{search}=7.3\times10^{-4}$ M$_{\odot}$ of the $\mathcal{M}_{c,injection} - \mathcal{M}_{c,trigger}$ distribution that is shown in Figure~\ref{fig:chirp-distrib}. Based on this, a random number is added to the injection chirp mass following a Normal distribution with parameters $\mu=0$ and $\sigma=\sigma_{search}$. Then \ba is used to compute the skymaps, and we test nine values of $\xi$ evenly spaced in $[0.7, 1]$ to which we add the baseline $\xi=0.83$ value. The exact values can be found in the legend of Figure~\ref{fig:pp-plot-all-blur}.
    

\begin{figure}
        \includegraphics[width=\columnwidth]{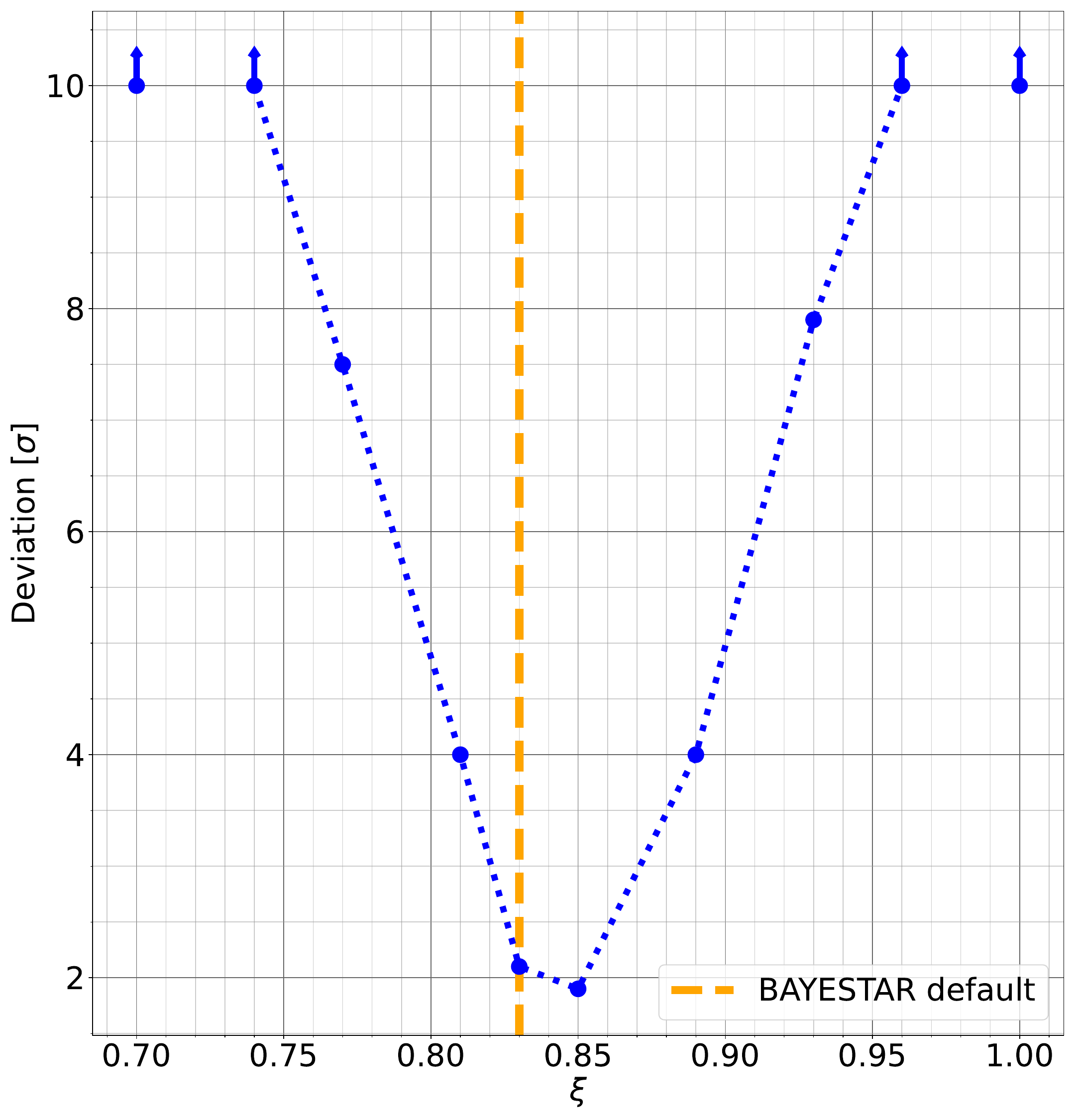}
        \caption{
        Optimisation of the $\xi$ parameter after randomizing the intrinsic parameters of the CBC template. The arrows show the values of $\xi$ for which the deviation from the null hypothesis of the KS test is larger than 10$\sigma$.}
        \label{fig:opt-all-blur}
    \end{figure}

    The PP-plots resulting from these simulations are shown in Figure~\ref{fig:pp-plot-all-blur}.
    Contrary to our previous results, here $\xi < 1$ is necessary to have a diagonal PP-plot; with $\xi=1$, the PP-plot sags below the diagonal, similar to the original observation made in \cite{bayestar}. This result suggests that $\xi$ effectively compensates for the difference between the true intrinsic parameters of the component objects, and the ``best'' values of those parameters identified by the matched filtering analysis.
    For a more precise inspection, we plot the deviation from the null hypothesis of the KS test as a function of $\xi$ in Figure~\ref{fig:opt-all-blur}. 
    This plot shows an optimum minimising the KS-test deviation for $\xi=0.85$, associated with a deviation of $1.9\sigma$. This value is close to the default setting $\xi=0.83$, showing that the simple test described in this Section allows us to reproduce almost entirely the effect observed with an online search.

\section{End-to-end simulations with \pcl}
\label{sec:live}

We now go one step beyond the simplified \texttt{pycbc\_make\_skymap} tool, and set up an end-to-end simulation where the injected signals are recovered by an actual \pcl analysis similar to what has been used during the O3 run. This gives us a more realistic configuration for optimizing the $\xi$ parameter than what is possible with \texttt{pycbc\_make\_skymap}.

As stated earlier, we expect two factors to dominate the discrepancy between the template and the source's intrinsic parameters: the noise realisation around the GW signal, which we cannot control, and the properties of the template bank used to run the search, which we are entirely free to choose.
The possibility that $\xi$ might be related to the template bank was already suggested in \cite{bayestar}.
Hence we perform analyses with several template banks to identify the parameters with the most prominent influence on $\xi$.
In particular, we start from the smallest and simplest bank possible and proceed towards the bank actually used in the O3 run.

\subsection{Simulated data and analysis method}
\label{sec:inj-set}

All the analyses presented in the following sections use the same dataset, which comprises 1000 BNS signals injected in simulated Gaussian noise, with the same distributions and parameters discussed in Section~\ref{sec:pop}. We choose simulated Gaussian noise as opposed to real data so as to have as much control as possible over the data, and avoid the practical complications of data gaps, data quality issues and actual astrophysical signals already present in real data. Based on the results of Section \ref{sec:real}, this choice is not expected to lead to a different conclusion compared to using injections in real data. The BNS signals are separated by $250$ s, excluding any overlap between the signals and ensuring the absence of a bias in the estimation of the noise ASD performed by \pcl. The total amount of simulated data represents about three days of data.

\pcl is run using a configuration equivalent to the one used for the online search during the O3 run \cite{pycbc_live_2}, except for the template bank, which changes for the various tests presented in the following subsections.

The GW candidates found by the search are associated with an injection when there is less than $40$ ms between the true and estimated coalescence times.
Then, we use \ba to localise the detected injections and produce the associated skymaps for constructing the PP plots. We use the same values of $\xi$ spread in $[0.7, 1]$, plus the baseline $0.83$ value, as in Figure~\ref{fig:pp-plot-all-blur} and generate skymaps for each of them. We generate a PP plot in a similar way as in Figure~\ref{fig:pp-plot-all-blur} and evaluate the deviation using the KS test for each value of $\xi$. This process allows us to identify the optimal $\xi$ value for which the deviation is minimal, in the same way as we did in Figure~\ref{fig:opt-all-blur}. 


\subsection{Template bank matching the injections}
\label{sec:bank-injection}

As we understand $\xi$ as a means to compensate for the difference between the template and true intrinsic parameters, we first test a ``minimal'' bank whose templates have the same intrinsic parameters as the injected signals.
Since the waveforms in this bank are extremely similar to the injected signals\footnote{Apart from minor differences between the \texttt{SpinTaylorT4} model used for the injections and the \texttt{TaylorF2} model used for the templates.}, we expect the optimal $\xi$ to be closer to $1$ than the baseline $0.83$. We cannot, however, exclude that a given injection could be best matched by a template corresponding to a different, nearby injection, leading to $\xi \ne 1$. Consequently, we perform a separate test where we explicitly select the \pcl candidates that exactly match the masses and spins of the injected signal.

\begin{figure}
    \includegraphics[width=\columnwidth]{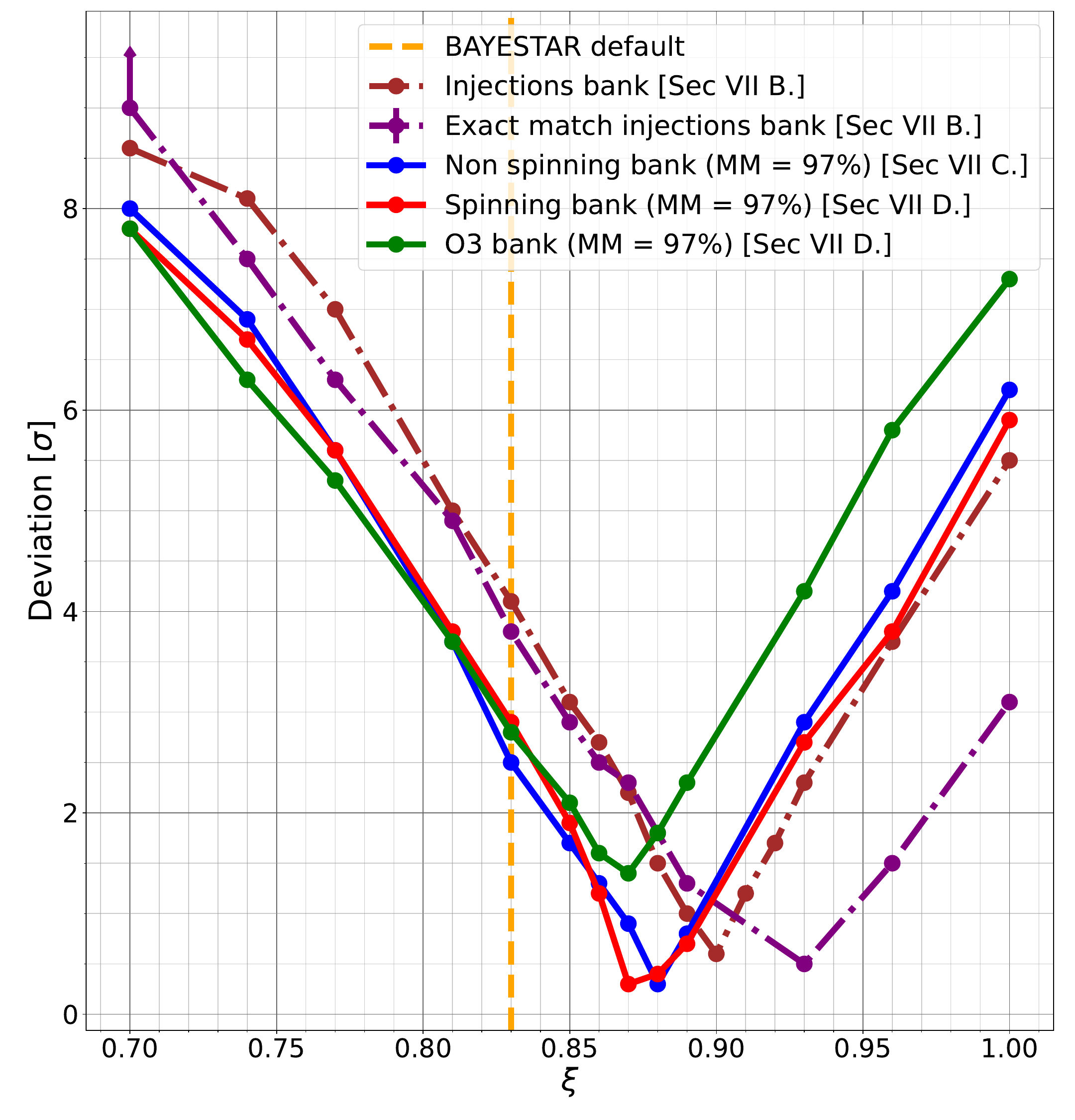}
    \caption{Optimisation of the $\xi$ parameter in \ba for \pcl. Each curve corresponds to a set of triggers produced by a \pcl analysis with a different template bank. The curves correspond to results presented in different subsections of Section~\ref{sec:live}, as detailed in the legend.
    The dashed-dotted curves correspond to the results obtained with the template banks based on the injection set described in Section~\ref{sec:bank-injection}. The purple curve corresponds to the results obtained with all the triggers produced by \pcl and the brown one to the subset of triggers exactly matching the injection they recovered. The blue curve corresponds to the template bank containing only BNS signals without spin and a minimal match of 97\%. The red curve corresponds to a BNS bank including nonzero spins and minimal match of 97\%. The green curve corresponds to the O3 bank. The vertical orange-dashed line corresponds to the baseline $\xi=0.83$ value implemented in \ba.}
    \label{fig:opt}
\end{figure}

About 500 injections are recovered by \pcl with this bank, which is similar to the tests presented in the later sections with more conventional banks. This excludes a pathological behaviour of the search with this bank. 350 triggers have an exact match with the injected signal.
Figure \ref{fig:opt} presents the optimisation curves obtained with this bank. The dashed-dotted brown line corresponds to the results for all the triggers. The one in purple corresponds to the results for the subset of triggers with an exact match between the injected and recovered templates.
For all the triggers mixed, the optimal value of $\xi$ is $0.90$ and for the exact match, $\xi$ is optimal at $0.93$. As expected, both values are indeed closer to $\xi = 1$ compared to Section \ref{sec:blurr}, and the exact-match result has the larger $\xi$. Note, however, that the optimal $\xi$ in the case of exactly-matching intrinsic parameters is still noticeably below $1$, which appears different from our previous results based on \texttt{pycbc\_make\_skymap}. This effect might originate from the differences between the \texttt{SpinTaylorT4} injections and the \texttt{TaylorF2} templates used by \pcl. The difference in approximant comes from a practical reason: \texttt{TaylorF2} is a frequency-domain model using the stationary phase approximation, while \texttt{SpinTaylorT4} is a time-domain model. The former is more suitable for frequency-domain matched-filtering, which PyCBC Live uses, while the latter is more suited for doing an injection and adding it to some GW data.

Together with the results of Section \ref{sec:blurr}, these results suggest that the template bank does play a role in the underestimation of the skymap uncertainty.

\subsection{Non-spinning banks of varying density}
\label{sec:non-spin-bank}

We now extend the search space of our bank to cover the entire range of component masses of our simulated signals, but set the component spins to zero for all templates.
We also test the hypothesis that using a sparser template bank may lead to an increase in the discrepancy between the true waveform parameters and the ones of the template reported by the search, thus driving the optimal $\xi$ to lower values than found in Section \ref{sec:bank-injection}.

Template banks are designed to have the smallest number of templates for maintaining computational efficiency without missing signals because of the mismatch between the model and the true signal. For a given vector of parameters $\Vec{\theta}$, the match between a template $h(\Vec{\theta})$ and a signal $h(\Vec{\theta}+\Vec{d\theta})$ is defined as \cite{PhysRevD.53.6749}:
        \begin{equation}
            M = \mathrm{max} \left( \frac{\langle h(\Vec{\theta})|h(\Vec{\theta} + d\Vec{\theta})\rangle}{\sqrt{\langle h(\Vec{\theta}) | h(\Vec{\theta}) \rangle \langle h(\Vec{\theta} + d\Vec{\theta}) | h(\Vec{\theta} + d\Vec{\theta}) \rangle}} \right)
        \end{equation}
where the maximization is carried out over time and phase shifts. A usual criterion to set the spacing between templates is to introduce a parameter called \emph{minimal match} ($MM$) such that $M \geq MM$ for any signal in the search space. $MM$ fixes the maximum acceptable fractional SNR loss incurred because of the sparseness of the bank. A commonly adopted value is $MM=97$\%, and this is also the value used for part of the template bank used by \pcl in O3.

We consider here three banks to test the impact of the sparseness on localization.
These banks are composed of low-mass, non-spinning CBC waveforms: a sparse bank with $MM=90$\%, one with the common $MM=97$\%  and a finer bank with $MM=98$\%.
These banks are generated using a geometrical placement described in \cite{hexa_template}, where the templates are placed in the parameter space on a hexagonal lattice. The generation is made with a \pycbc tool called \texttt{pycbc\_geom\_nonspinbank}.
The templates of these banks are post-Newtonian waveforms at the $3.5$ order and have both component masses in the $[1,12]$M$_{\odot}$ range to safely cover our BNS injections.
We then apply a cut on the chirp mass at $2.7$M$_{\odot}$, so as to reduce the number of templates as much as possible, while still covering the BNS injections and avoiding border effects potentially arising from a cut on component masses. We stress explicitly that, contrary to the bank used in Section \ref{sec:bank-injection}, the banks here are placed independently of the injections, and do not include templates with the true injection parameters.

\begin{figure}
            \includegraphics[width=\columnwidth]{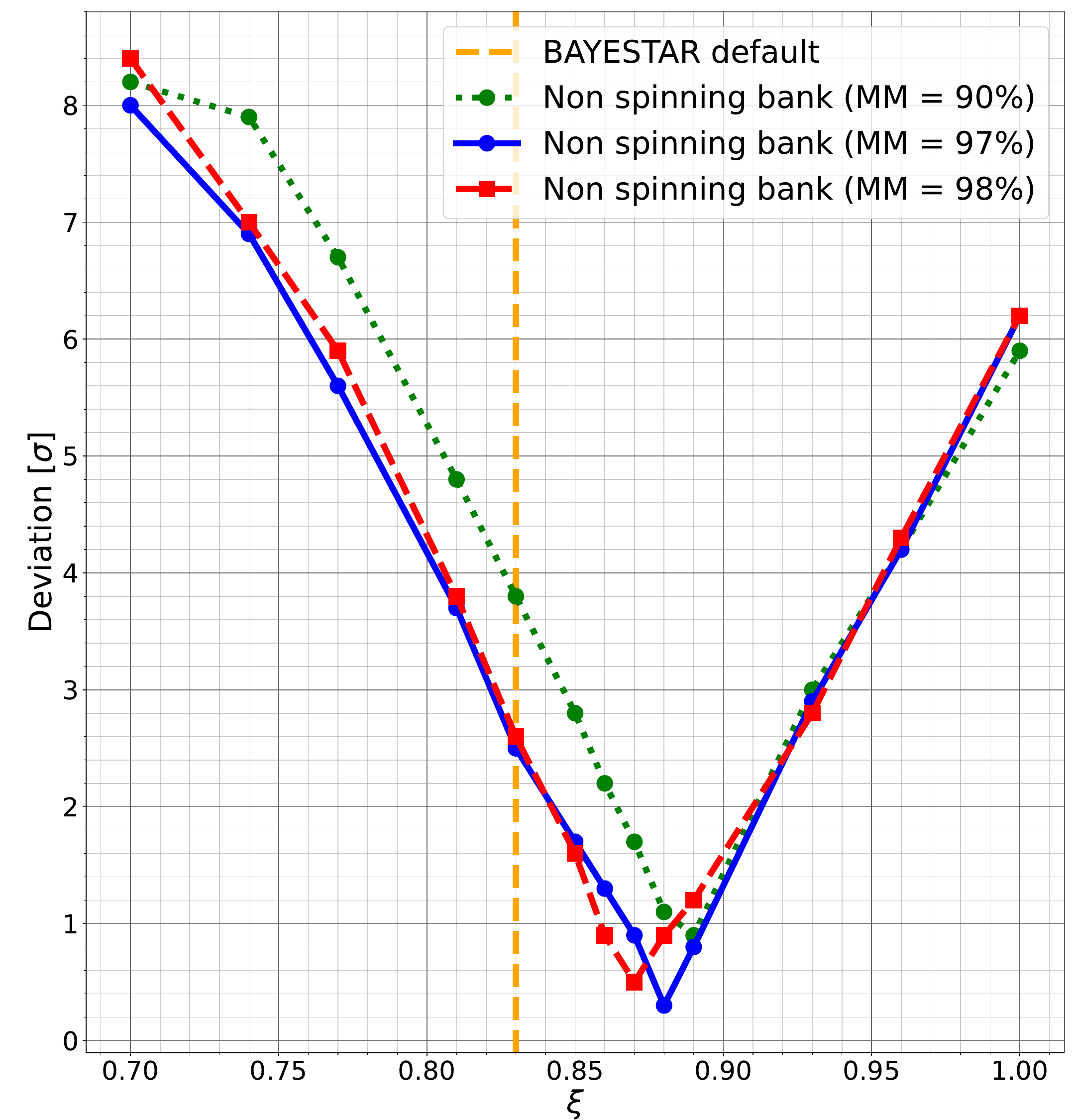}
            \caption{Optimisation curves for low-mass, zero-spin template banks with minimal matches of 90\% (green-dotted), 97\% (blue-solid) and 98\% (red-dashed).}
            \label{fig:opt-sparse}
\end{figure}

Figure \ref{fig:opt-sparse} shows the results for the three banks, and the blue curve of Figure \ref{fig:opt} shows the results for $MM=97$\% to compare to other template banks of this Section.
Remarkably, the optimal $\xi$ value \emph{decreases} when the $MM$ increases, although this effect is rather small, as several values of $\xi$ produce deviations below $1\sigma$, indicating acceptable performances.
We also observe that the optimal $\xi$ is further reduced with respect to the case where the exact injections are used as a template bank.
We conclude that the density of templates appears to have a small but detectable effect on the optimal choice of $\xi$, as suggested in \cite{bayestar}.
The effect goes into a somewhat counterintuitive direction, as denser banks appear to require a stronger correction.
It also appears that the amount of explored parameter space around each signal may also have a small effect on the optimal $\xi$.

\subsection{Adding aligned spins to the search space}
\label{sec:spin-bank}

As the next step, we include spins in the search parameter space.
The mass ranges are the same as in the previous section, but we also include aligned spins with magnitudes between $0$ and $0.05$ for neutron stars and between $0$ and $0.998$ for black holes.
We use a minimal match of 97\%, for this is the usual value for the actual search banks. We stress again that, contrary to the bank used in Section \ref{sec:bank-injection}, the bank here is placed independently of the injections, and does not include templates with the true injection parameters.
The curve in red in Figure \ref{fig:opt} shows the results for localising the triggers found with this bank.

The optimal value of $\xi$ is $0.87$, slightly smaller than the zero-spin bank with the same $MM$ as described in the previous section.
This result goes in the direction consistent with the hypothesis that $\xi$ is somehow related to the extent (or dimensionality) of the search space around each signal.

\subsection{O3 bank}
\label{sec:ff-live}

As a last check, we use the same bank used by \pcl during the O3 run on the simulated data. The bank design is detailed in \cite{pycbc_live_2, 2017arXiv170501845D, 2017PhRvD..95j4045R, PhysRevD.99.024048}, and the parameter space explored by this bank includes BNS, BBH and NSBH. For BNS and NSBH binaries, the neutron stars have masses between 1 and 2 $M_{\odot}$ and aligned spins with magnitude up to $0.05$. These boundaries account for the masses and largest spins observed in pulsars \cite{Lorimer2008}. For black holes, the spin magnitudes are up to $0.998$, accounting for the spins observed in X-ray binaries \cite{McClintock2014}. The total mass of the binaries goes up to $100 M_{\odot}$ for equal mass, weakly spinning templates and up to $500 M_{\odot}$ for templates with high mass ratio, and high aligned spins. We stress once more that, contrary to the bank used in Section \ref{sec:bank-injection}, the bank used here is placed independently of the injections, and does not include templates with the true injection parameters.

The results for the O3 bank are visible as the green curve in Figure~\ref{fig:opt} and show an optimal $\xi$ of $0.87$, slightly higher than the baseline $0.83$ value used by default in \ba. 
This optimal value is compatible with the previous bank with $MM=97$\% and nonzero spins.
The difference with the default $\xi$ suggests that the sky localizations for \pcl BNS events are, on average, slightly broader than they should be.
As such, $\xi$ should be slightly increased when localising BNS candidates produced by \pcl, particularly considering that the deviation at the baseline $\xi$ approaches the $3\sigma$ level.

\section{Interpretation}
\label{sec:inter}

The various tests carried out in the previous sections indicate that the optimal value of $\xi$ is affected in varying amounts by a variety of possible effects. First, we have evidence from Section \ref{sec:network-config} that the optimal $\xi$ seems to depend on the composition of the detector network, i.e.~on the number of detectors and their relative sensitivities. In particular, smaller or more geometrically-degenerate networks appear to lead to an optimal $\xi$ closer to $1$.
A second conclusion we can make is that real detector noise, which is non-stationary and non-Gaussian, does not appear to play a significant role in terms of $\xi$. This is somewhat consistent with the findings in \cite{Macas2022}, especially for BNS mergers, though that study focused on very specific cases. On the other hand, an important driver of the optimal $\xi$ appears to be related to the inevitable discrepancy between the true intrinsic parameters of a CBC signal, and those of the templates identified by the matched filtering search that produced the candidate event. This was in fact suggested as a possibility in \cite{bayestar}. In particular, if we pretend that the search will exactly identify the true intrinsic parameters, then the $\xi$ correction is not necessary. However, as soon as we allow a deviation, we immediately observe that a value $\xi \sim 0.87$ becomes necessary to properly compute the uncertainty in the sky localization. The optimal value depends on how exactly the deviation takes place: if we allow only a ``minimal'' deviation, corresponding to a template bank that only explores the set of simulated signals, then the optimal $\xi$ is closer to $1$. As we introduce more and more templates in the vicinity of the true values of each signal (either by increasing the minimal match of the bank, or by including additional dimensions in the search space, such as spins) the optimal value tends to further decrease, albeit by a smaller amount. Perhaps counter-intuitively, we find that making the bank denser leads to a slightly \emph{smaller} optimal $\xi$. Therefore, it appears that the bias is originating from the ``freedom'' that the search template has to slightly deviate from the true parameters.

Based on these observations, we propose the following argument.
\ba makes the assumption that the uncertainties in the spatial location parameters are uncorrelated with those in the intrinsic parameters.
This assumption is certainly appropriate, and it introduces no systematic shift of the posterior distribution of the spatial parameters.
However, we argue that it is implicitly acting as a delta-prior on the intrinsic parameters, effectively assuming that the search template is exactly matching the true intrinsic parameters.
If this was true, then the spread of \ba's posterior distribution for the spatial position would match what it would be if \ba carried out a full parameter estimation of the entire vector of unknown signal parameters, followed by a marginalization over the intrinsic parameters.
However, the matched-filter search returns effectively one sample from the peak of the likelihood function over the intrinsic parameters, and this sample will in general not be at the true values of the parameters, and for technical reasons may not even be at the maximum of the likelihood (see e.g.~the discussion in \cite{pycbc_live_2} for PyCBC Live).
Therefore, the spread of \ba's posterior distribution for the spatial position will be slightly underestimated with respect to a full parameter estimation followed by marginalization.
The $\xi$ factor is then used to compensate this underestimation by introducing an additional spread in the spatial posterior distribution.
Its optimal value is therefore related to how likely the single sample of the intrinsic parameters is to be at a certain distance from the true intrinsic parameters, and this ultimately depends on how many templates are ``available'' around the true parameters.

To conclude, we argue that $\xi$ has to be set in a way that depends on the specific properties of the template bank used by the search, namely its density (or minimal match), its dimensionality, and the extent of the search space explored by it.
With more work it might be possible, for instance, to relate $\xi$ to the average match between the true waveforms and the templates reported by the search.
We consider our simulations as an initial step towards understanding this dependence.

\section{Conclusion}
\label{sec:conclusion}

In this work, we consider GW candidate events produced by the \pcl detection pipeline and present an analysis for optimizing the low-latency sky localization of those events with the \ba algorithm. The latter provides skymaps within seconds after detecting a GW signal with online searches. This information is crucial for multi-messenger astronomy and electromagnetic follow-up, making the low-latency localization a critical part of the follow-up. We evaluate the accuracy of \ba by simulating a population of CBC signals, recovering and localizing them under a variety of assumptions, and checking the consistency of the resulting sky localization using a common test based on the PP plot, which is expected to be diagonal for correctly estimated uncertainties.

\ba has a tuning parameter $\xi$ that allows one to make the sky localization self-consistent. It acts as a scaling factor for the SNR reported by the search pipeline before the Bayesian inference. Originally this factor had been tuned based on a specific set of assumptions for passing the PP-plot test, resulting in a default value of $0.83$ \cite{bayestar}. We first show that an idealized condition exists where such a correction is not necessary (equivalent to setting $\xi=1$), namely when the true intrinsic parameters of a signal exactly match those reported by the search pipeline. We also show that the correction ($\xi < 1$) becomes necessary as soon as there is even a minor mismatch between the true intrinsic parameters of a signal and those reported by the search pipeline, which of course is always the case in practice.

Focusing on BNS mergers, which are the most promising sources of EM counterparts, we use a series of end-to-end tests with \pcl in order to identify the search parameters influencing $\xi$. Starting with a template bank for searching for GW signal that closely matches the simulated signal, we observe that the optimal value for $\xi$ is closer to $1$ compared to the baseline value. We also use three template banks with only BNS signals but different minimal matches, a parameter controlling the bank sparseness. Based on our results, we conclude that the optimal $\xi$ depends to a small extent both on the sparseness of the template bank and on its dimensionality, consistently with what was suggested in \cite{bayestar}. Finally, we test the localization with the template bank used for the O3 analysis, and we find that the optimal value is $\xi \sim 0.87$, i.e.~closer to the default than our previous tests, but slightly larger.
This suggests that a dedicated tuning campaign of $\xi$ for \pcl, and more generally for any search using template banks significantly different than what was used in \cite{bayestar} for \gstlal, might improve the consistency of \ba's sky localizations.
We also find that the consistency of the localization for BNS systems does not appear to depend significantly on the quality of the data, i.e.~on differences between real detector noise and stationary Gaussian noise. This is somewhat consistent with the findings of \cite{Macas2022}. We have evidence that the optimal $\xi$ might depend on the characteristics of the detector network. In particular, networks that are less capable of performing precise localizations might require a smaller correction. Exploring this dependence precisely is left to future work, which might be important as KAGRA and LIGO-India progress towards sensitivities comparable to LIGO and Virgo.

Tests similar to ours have been carried out recently in \cite{Chaudhary2023}, using a dedicated end-to-end mock data challenge in low-latency, but without exploring the $\xi$ parameter in great detail.
The resulting PP plots appear to indicate consistent sky localizations for all search pipelines currently operating in the O4 run.
This is both reassuring, and perhaps not surprising considering that the setup of that simulation, and in particular the properties of the simulated signals, are quite different from our choices here.

In this work, we have only explored the consistency of the posterior distribution for the two-dimensional position on the sky, focusing on the main needs of optical telescopes.
Similar investigations should be carried out in the future for the luminosity distance posterior, also provided by \ba, or more generally for the joint three-dimensional spatial distribution.
Another avenue worth investigating is the effect of the post-discovery SNR maximization done in O3 by \pcl, described in \cite{pycbc_live_2}. Since this method should lead to an improved match between the true source parameters and the selected template, it should in principle lead to an optimal $\xi$ closer to $1$. Consequently, a simulation campaign testing the influence of this method should be made in the future.

\acknowledgements

We thank Leo Singer for useful discussion and help with the \ba code, and Thomas Dent, Michael Coughlin, Ed Porter, Colm Talbot and Jolien Creighton for useful comments on the study and draft.  We are also grateful to the anonymous referee for providing valuable comments.
TD additionally thanks FP for the kind hospitality at the Sapienza University of Rome during part of this work.
SA thanks the CSI Recherche University Côte d'Azur for their financial support. FP acknowledges support from the ICSC - Centro Nazionale di Ricerca in High Performance Computing, Big Data and Quantum Computing, funded by the European Union - NextGenerationEU and support from the Italian Ministry of University and Research (MUR) Progetti di ricerca di Rilevante Interesse Nazionale (PRIN) Bando 2022 - grant 20228TLHPE - CUP I53D23000630006.

Part of our simulations utilized the Virtual Data cloud computing system at IJCLab.
We thank Michel Jouvin and Gerard Marchal-Duval for their prompt support and advice about this system. The authors are also grateful for computational resources provided by the LIGO Laboratory and supported by National Science Foundation Grants PHY-0757058 and PHY-0823459. 

This manuscript has LIGO document number LIGO-P2300429.

\bibliographystyle{apsrev4-2}
\bibliography{references}

\end{document}